\documentstyle[psfig,epsf,amssymb]{elsart}
\begin{document}
\begin{frontmatter}
\title{Isotropic phase squeezing and the arrow of time}
\author{Giacomo M. D'Ariano, Chiara Macchiavello,}
\author{Paolo Perinotti and Massimiliano F. Sacchi}
\address{{\sc Theoretical Quantum Optics Group}\\
Dipartimento di Fisica 'Alessandro Volta', Universit\`a di Pavia \\
Istituto Nazionale di Fisica della Materia -- Unit\`a di Pavia \\
via Bassi 6, I-27100 Pavia, ITALY}
\date{\today}\maketitle
\begin{abstract}
We prove that isotropic squeezing of the phase is equivalent to
reversing the arrow of time.
\end{abstract}
\end{frontmatter}
The concept of ``squeezing'' appeared in the literature in the early
70's \cite{stl,hpy} and was extensively studied in order to improve
the capacity of quantum information channels and the sensitivity in
interferometric measurements \cite{tomb}. Since then squeezing has
been a very popular word in quantum optics.  By ``squeezing'' one
refers to a physical process where the uncertainty of an observable is
reduced at the expense of increasing the uncertainty of the conjugated
observable, according to the Heisenberg inequalities (for extensive
reviews see for example Ref. \cite{spe}). In quantum optics quadrature
squeezing, namely the squeezing of the probability distribution of the
observable $a_{\phi}=(a^{\dag}e^{i\phi}+ae^{-i\phi})/2$---$a$ and
$a^{\dag}$ denoting the annihilation and creation operators of a given
radiation mode---has been achieved experimentally, giving rise to a
number of interesting properties, such as phase-sensitive
amplification and antibunching \cite{hpy,spe,yam,iss}.  More recently,
the density operator of squeezed states has been measured by optical
homodyne tomography \cite{schiller}.  \par
\begin{figure}
\epsfxsize=0.45\hsize\leavevmode\epsffile{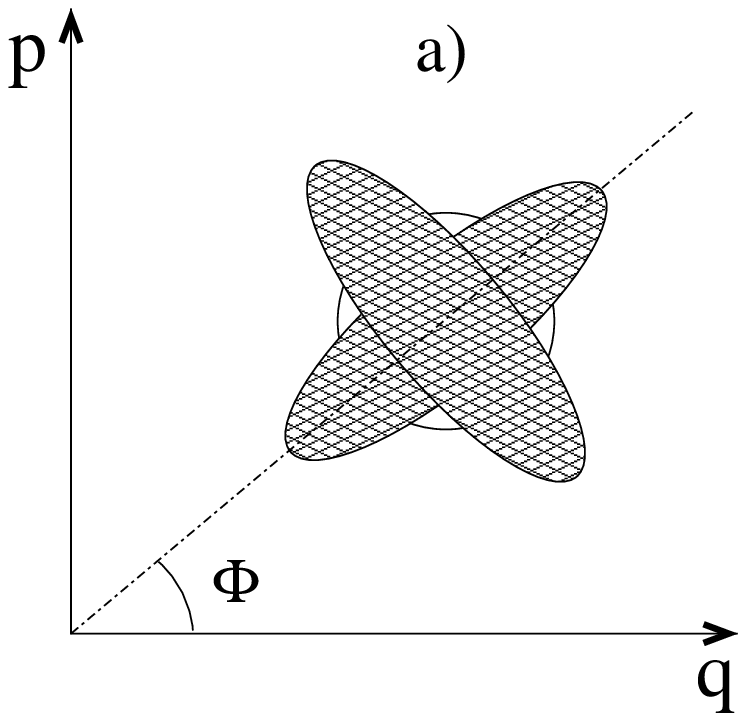}\hspace{16pt}
\epsfxsize=0.45\hsize\leavevmode\epsffile{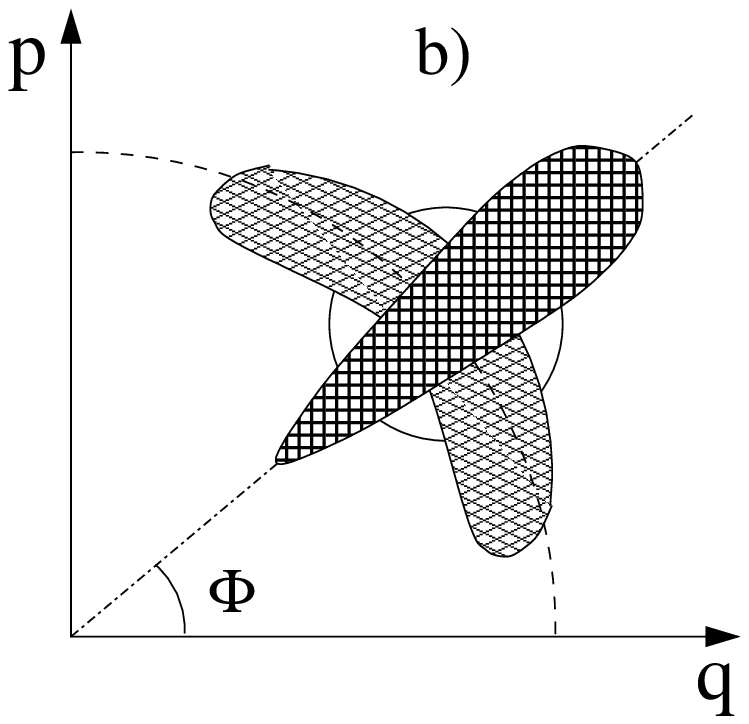}
\vskip 3truecm
\caption{Phase-space representation of squeezing: (a) conventional
squeezing of two conjugated quadratures; (b) squeezing in phase and
photon number.}
\label{f:fig1}
\end{figure}

 Conjugated quadratures,
i.e. quadratures relative to phases $\phi$ and $\phi +\pi /2$, are
generalizations of the couple of observables position-momentum. Thus,
we can view quadrature squeezing of radiation states as a narrowing
process of the probability distribution in the phase space which
occurs in a definite direction, corresponding to the phase of the
squeezed quadrature [see Fig. \ref{f:fig1}(a)]. 
Classically, one can imagine a similar process in polar
coordinates [Fig. \ref{f:fig1}(b)], where the radial probability
distribution is squeezed, while the phase is spread, or viceversa. In
the phase space the squared radius corresponds to the total energy of
the harmonic oscillator, which is proportional to the photon number
operator $N=a^\dag a$ for a single-mode radiation field. Number
squeezing narrows the photon number distribution, with the possibility
of achieving sub-Poissonian statistics $\langle\Delta N^2\rangle
<\langle N \rangle$ in photon counting \cite{yaim}.  This process has
been investigated extensively and can be experimentally achieved by
means of self-phase modulation in Kerr media \cite{banana}.  \par The
inverse process, namely phase squeezing, is the subject of the present
Letter. We will consider isotropic phase squeezing, namely squeezing
of the phase probability distribution independently of the mean value
of the phase.  Such a process corresponds to noise reduction in the
measurement of phase, and it would lead to important results for
communications and measurements, such as improved sensitivity of
interferometric schemes and the achievement of the capacity of quantum
communications based on phase coding.  \par In the following we will
prove that isotropic phase squeezing cannot be realized because it
would correspond to reversing the arrow of time. The arrow of time is
statistically defined by the direction of the irreversible dynamics of
open systems \cite{arrow}. In quantum-mechanical terms, it is
associated to a loss of coherence of the quantum state, e.g.
dephasing mechanism of the laser light, which corresponds to a random
walk on the phase space \cite{randomlaser}. We will prove that any
dynamical process that isotropically reduces the phase uncertainty can
be described only in terms of a ``time-reversed dissipative
equation''.  \par In the literature the Heisenberg-like heuristic
inequality $\Delta N\Delta\phi\geq 1$ for the couple number-phase is
often reported. However, its meaning is only semiclassical, since the
quantum phase does not correspond to any self-adjoint operator
\cite{sg,shsh,ban}. Therefore, in order to investigate isotropic phase
squeezing, we have first to introduce the concepts of phase
measurement and phase probability distribution in a rigorous way.

\par The quantum-mechanical definition of the phase is well assessed
in the framework of quantum estimation theory \cite{stima1,stima2}. In
this context the phase of a quantum state is defined by the shift
$\phi $ generated by any operator $F$ with discrete spectrum. For
example $F=a^{\dag }a$ for the harmonic oscillator, and $F=\sigma
_z/2$ for a two-level system, $\sigma _z$ being the customary Pauli
operator.  \par Quantum estimation theory provides a general
description of quantum statistics in terms of POVM's (positive
operator-valued measures) and seeks the optimal POVM to estimate one
or more parameters of a quantum system on the basis of a cost function
which assesses the cost of errors in the estimates.  For phase
estimation, the optimal POVM for pure states $|\psi \rangle $ with
coefficients $\psi _n =|\psi_n |e^{i\chi_n}\not=0$ on the basis
$|n\rangle$ of $F$ eigenvectors is given by
\begin{eqnarray}
d\mu (\phi)=
\frac{d\phi }{2\pi} |e(\phi)\rangle \langle
e(\phi)|
\label{pomopt}
\end{eqnarray}
for the class of Holevo's cost functions---a large class including the
maximum likelihood criterion, the $2\pi$-periodicized variance, and
the fidelity optimization.  In Eq. (\ref{pomopt}) $|e(\phi)\rangle $
denotes the (Dirac) normalizable vector
\begin{eqnarray}
|e(\phi)\rangle =\sum_{n\in S} e^{i(n\phi-\chi_n)} |n\rangle \;,
\label{sg}
\end{eqnarray}
where $S$ is the spectrum of $ F$.  In Ref. \cite{dms} the solution
given in Eqs. (\ref{pomopt}-\ref{sg}) has also been proved for {\em
phase-pure} states, namely for states described by a density operator
$\rho$ satisfying the condition
\begin{eqnarray}
\rho _{nm}\equiv\langle n|\hat\rho |m \rangle = |\rho
_{nm}|\,e^{i(\chi _n -\chi_m)}\;,\end{eqnarray} and for a
nondegenerate phase-shift generator $F$ \cite{notabene}.  For states
that are not of this kind, there is no available method in the
literature to obtain the optimal POVM, and thus the concept itself of
phase does not have a well defined meaning.  \par The phase
probability distribution $dp(\phi)$ of a quantum state is evaluated by
means of the optimal POVM in Eq. (\ref{pomopt}) through the Born's
rule $dp(\phi)={\rm Tr}[\rho d\mu(\phi)]$.  The phase uncertainty
$\Delta\phi^2$ of the state can then be calculated.  However, notice
that for periodic distributions the customary r.m.s. deviation depends
on the chosen window of integration.  Definitions of phase uncertainty
that do not depend on the interval of integration given in the
literature are monotonic increasing functions $f$ of some average cost
of the Holevo's class, namely
\begin{eqnarray}
\delta\phi =f( \langle  C \rangle )\;,\label{D1}
\end{eqnarray}
where $ C$ represents the cost operator
\begin{eqnarray}
 C=c_0 -\sum
_{n=1}^{\infty} c_n\, (e_+^n+e_-^n)\,,\qquad c_n\geq 0\,,\quad\forall
n\geq 1\,,\;\label{D2}
\end{eqnarray}
and $\langle ... \rangle $ represents the quantum ensemble average.
In Eq. (\ref{D2}) we introduced the following notation 
\begin{eqnarray}
e_+=\sum_{n\in S}e^{i(\chi_{n+1}-\chi_n )}|n+1\rangle \langle n|\quad
\hbox{and } e_-=(e_+)^\dag \;.\end{eqnarray} Typical examples of
functions of this kind are the reciprocal peak likelihood and the
phase deviation $2(1-|\langle e_+ \rangle |^2)$.  The former
corresponds to $f(x)=-1/x$ for the cost operator $C$ with all $c_n=1
$; the latter corresponds to $f=2\big[1-(1/4)x^2\big]$ with $c_1 =1$
and $c_n =0$ $\forall n\not=1$ (for phase-pure states $\langle e_+
\rangle$ is a real positive quantity, so $\langle e_- \rangle=\langle
e_+ \rangle$ and $-\langle C\rangle^2 /4=-\vert \langle e_+
\rangle\vert ^2$, with $\langle C\rangle\leq 0$).  \par Now we
introduce the concept of isotropic phase squeezing. For the
e.m. field, we remind that ordinary quadrature squeezing is effective
in reducing the phase uncertainty of a quantum state provided that the
average value of the phase is known {\em a priori}.  As mentioned
above, isotropic phase squeezing should reduce the phase uncertainty
of the state $\rho $ independently of the initial mean phase. In
mathematical terms, this condition corresponds to a linear map $\Gamma
$ that is covariant for the rotation group generated by the operator $
F$, namely
\begin{eqnarray}
\Gamma (e^{i F\phi}\,\rho \,e^{-i F\phi})=
e^{i F\phi}\,\Gamma (\rho )\,e^{-i F\phi}
\;.\label{covgamma}
\end{eqnarray}
A physically realizable linear map $\Gamma $ corresponds to a completely
positive (CP) map for density operators that can be
written in the Lindblad form 
\begin{eqnarray}
\frac{\partial \rho}{\partial t }=\sum_n L[V_n] \rho \;,\label{meq}
\end{eqnarray}
where $L[O]\rho \equiv O\rho O^\dag -\frac 12 (O^\dag O \rho +\rho
O^\dag O )$ denotes the Lindblad superoperator \cite{lind}.  We do not
take into account the customary Hamiltonian term $-i[H,\rho ]$ in the
master equation (\ref{meq}), because for the covariance condition one
has $[H,F]=0$, and hence the optimal POVM is simply rotated, with the
result that the phase uncertainty is not affected by the corresponding
unitary evolution (such Hamiltonian term can be equivalently applied
in one step before the evolution (\ref{meq}), and preserves the phase
purity of the state).  \par The covariance condition restricts the
general form of Eq. (\ref{meq}) to the expression \cite{holevo}
\begin{eqnarray}
\frac{\partial \rho}{\partial t }=\sum_{m=-\infty}^{+\infty} 
\sum_j L[B_{m,j}] \rho \;,\label{meq2}
\end{eqnarray}
where
\begin{eqnarray}
&B_{mj}&=g_{m,j}( F)\, e_+^m \,,\qquad m\geq 0 \nonumber 
\\&B_{mj}&=h_{|m|,j}( F)\,e_-^{|m|} \,,\qquad m<0\;.\label{2exp}
\end{eqnarray}
In the following we will focus our attention on the case of a
single-mode radiation field, hence we take $ F=a^{\dag }a$ and the
spectrum $S=\mathbb N$.  We postpone the discussion of the generality
of our result at the end of the paper.  \par We consider the class of
phase-pure states as initial states for the master equation
(\ref{meq2}), since for other kinds of states the phase measurement is
not well defined, as mentioned before.  \par We are now in position to
prove the main result of this paper, namely that isotropic phase
squeezing is equivalent to reversing the arrow of time. According to
Eq. (\ref{D1}) the time derivative of the phase uncertainty
$\delta\phi$ has the same sign as the derivative of the average cost
$\langle C \rangle$, which is obtained from Eq. (\ref{meq2}) as follows
\begin{eqnarray}
\frac{\partial \langle  C\rangle }{\partial t }=
-2 \mbox{Re}\frac{\partial }{\partial t }\sum_{k=1}^{\infty}
c_k \langle e_+^k\rangle \;.\label{derc}
\end{eqnarray}
A straightforward calculation gives the following contribution for 
the $k$-th term in the sum of Eq. (\ref{derc})
\begin{eqnarray}
&&-2 \mbox{Re}\frac{\partial }{\partial t }
\langle e_+^k\rangle = \label{nth}\\&&  
\sum_{m=0}^{\infty}\sum_j \sum _{l=0}^{\infty }|\rho _{l,l+k}| |
g_{m,j}(m+l)-g_{m,j}(m+l+k)|^2 \nonumber \\&&+
\sum _{m=1}^{\infty }\sum_j \sum _{l=m}^{\infty }|\rho _{l,l+k}| |
h_{m,j}(l-m)-h_{m,j}(l-m+k)|^2 \nonumber \\&&  +
\sum _{m=1}^{\infty }\sum_j \sum _{l=\max(0,m-k)}^{m-1}|\rho _{l,l+k}| 
|h_{m,j}(l-m+k)|^2 
\geq 0 \; ,\nonumber 
\end{eqnarray}
which is manifestly non negative for all values of $k$. Hence,
according to Eqs. (\ref{derc}) and (\ref{nth}), the average cost as
well as the phase uncertainty increase versus time. The only
possibility to achieve isotropic phase squeezing is then to have a
minus sign in front of the master equation (\ref{meq}), which means to
reverse the arrow of time.  \par Our proof rules out also the
possibility of isotropic squeezing through a phase measurement
followed by a feedback quadrature squeezing. In fact, such a kind of
process is described by a CP map as well, and one can explicitly show
that the phase uncertainty in the measurement eventually leads to an
overall phase diffusion.  \par The same conclusions regarding the
derivatives of $\langle e_+^k\rangle$ hold also for the cases of
unbounded spectrum $S=\mathbb Z$ and bounded spectrum $S={\mathbb
Z}_q$ for a nondegenerate phase shift operator: also in these cases
the phase uncertainty can decrease for any input state only if we
reverse the arrow of time.  For $S={\mathbb Z}_q$ all series in
Eq. (\ref{nth}) are bounded and boundary terms appear in addition to
the third one.  For $S={\mathbb Z}$ Eq. (\ref{nth}) rewrites:
\begin{eqnarray}
&&-2 \mbox{Re}\frac{\partial }{\partial t }
\langle e_+^k\rangle = 
\label{nth'}
\\&&  
\sum_{m=0}^{\infty}\sum_j \sum _{l=0}^{\infty }|\rho _{l,l+k}| |
g_{m,j}(m+l)-g_{m,j}(m+l+k)|^2 \nonumber \\&&+
\sum_{m=1}^{\infty}\sum_j \sum _{l=1}^{\infty }|\rho _{l,l+k}| |
h_{m,j}(l-m)-h_{m,j}(l-m+k)|^2 \geq 0 \; .\nonumber 
\end{eqnarray}
\par In the general case a phase-covariant master equation does not
evolve a phase-pure state into a phase-pure state, hence it may happen
in principle that, after a finite time interval in which the
phase-purity is lost, phase-purity is then recovered at the end with
an overall decrease of $\delta\phi$.  However, one cannot follow the
evolution of $\delta\phi$ for finite time intervals if the definition
itself of the phase is lost during the time evolution.  A
phase-covariant master equation (\ref{meq2}) preserves phase-purity if
and only if $\arg(g_{m,j} (F))=\varphi_{m,j}$ and 
$\arg(h_{m,j} (F))=\theta_{m,j}$ independent on $F$, and one can
always choose $\varphi_{m,j}=\theta_{m,j}=0$ identically, due to the
bilinear form of the Lindblad superoperator. \par In the
case of degenerate $F$, we can find the optimal POVM for pure states
$\vert\psi\rangle\langle\psi\vert$ as follows \cite{dms}: we select a
vector $\vert n\rangle_{\parallel}$ for each degenerate eigenspace
${\cal H}_n$ corresponding to the eigenvalue $n$, such that $\vert
n\rangle_{\parallel}$ is parallel to the projection of
$\vert\psi\rangle$ on ${\cal H}_n$.  So the Hilbert space ${\cal H}$
can be represented as ${\cal H}_{\parallel}\oplus{\cal H}_{\perp}$,
where ${\cal H}_{\parallel}$ is the Hilbert space spanned by the
vectors $\vert n\rangle_{\parallel}$ and ${\cal H}_{\perp}$ its
orthogonal completion. Since $\vert\psi\rangle$ has null component in
${\cal H}_{\perp}$ the estimation problem reduces to a nondegenerate
one in the Hilbert space ${\cal H}_{\parallel}$, and the optimal POVM
is given by $d\mu (\phi)=d\mu_{\parallel}(\phi)\oplus
d\mu_{\perp}(\phi)$, where $d\mu_{\parallel}$ is the optimal POVM for
the nondegenerate estimation problem in ${\cal H}_{\parallel}$, while
$d\mu_{\perp}(\phi)$ is an arbitrary POVM in ${\cal H}_{\perp}$.  It
is clear that the POVM obtained in this way is optimal also for
phase-pure states that are mixtures of pure states all with the same
${\cal H}_{\parallel}$.  The most general phase-covariant master
equation is again of the form in Eqs. (\ref{meq}) and (\ref{2exp}),
however, now there are infinitely many possible $B_{m,j}$ for fixed 
$m,j$ corresponding to different operators $e_+$ which shift the eigenvalue
of $F$ while spreading the state in the whole $\cal H$ from ${\cal
H}_{\parallel}$ in all possible ways.  Therefore, a phase-covariant
master equation does not keep the original state in ${\cal
H}_{\parallel}$, apart from the case where one considers operators
$B_{m,j}$ defined in terms of $e_+$ only of the form
\begin{eqnarray}
e_+ =\sum_n e^{i(\chi_{n+1}-\chi_n)}
|n+1\rangle_{\parallel\parallel}\langle n|\;.\end{eqnarray} In the
general case, however, a reduction of the phase uncertainty is
possible in principle, due to the arbitrariness introduced by
$d\mu_{\perp}(\phi)$ in the definition of phase in the degenerate
case, the time derivative of $dp(\phi)={\rm Tr}[\rho d\mu
_{\parallel}(\phi)]$ generally depending on $d\mu_{\perp}(\phi)$.
\par We now focus attention back to the case of nondegenerate $F$.
Looking at Eq. (\ref{nth}), one can see that it is possible to
make all terms vanishing, getting a null derivative for the average
cost.  This is actually possible only for unbounded spectra as
$S={\mathbb N}$ and $S={\mathbb Z}$, where one can find a master
equation that preserves the phase uncertainty for any quantum
state. For $S=\mathbb N$, one has the following conditions on the
coefficients of Eq. (\ref{nth}): $g_{m,j}(F)=g_{m,j}$ constant and
$h_{m,j}(F)=0$. Upon defining $u_m=\sum _j |g_{m,j}|^2$ one has 
\begin{eqnarray}
\frac{\partial \rho}{\partial t }=\sum_{m=1}^\infty 
u_m (e_+^m \rho e_-^m-\rho)\;.
\label{pp}
\end{eqnarray}
In the case $S=\mathbb Z$ one has more generally $h_{m,j}(F)=h_{m,j}$
constant, and introducing $v_m=\sum _j |h_{m,j}|^2$, the
phase-uncertainty preserving master equation takes the form
\begin{eqnarray}
\frac{\partial \rho}{\partial t }=
\sum_{m=1}^\infty u_m (e_+^m \rho e_-^m-\rho)
+v_m (e_-^m \rho e_+^m-\rho)\;.
\label{pp2}
\end{eqnarray}
The master equations (\ref{pp}) and (\ref{pp2}) are very interesting,
since they represent a counterexample to the customary identification
of ``decoherence'' and ``dephasing''. The study of physical
realizations of Eqs. (\ref{pp}) and (\ref{pp2}) could provide insight
in the understanding of decoherence and relaxation phenomena.  \par In
conclusion, we have shown that isotropic squeezing of the phase is
equivalent to reversing the arrow of time. This result is very
general, as it holds for any definition of phase with nondegenerate
shift operator, for any definition of phase-uncertainty in the
Holevo's class, and for any initial phase-pure state.  In this way we
have related the concept of phase to the arrow of time statistically
defined by the evolution of open quantum systems, thus enforcing the
link between phase and time \cite{peres}.  \par This work is supported
by the Italian Ministero dell'Universit\`a e della Ricerca Scientifica
e Tecnologica under the program {\em Amplificazione e rivelazione di
radiazione quantistica}, and by the INFM PAIS 1999. 

\begin{thebibliography}{99}
\bibitem{stl} D. Stoler, Phys. Rev. D {\bf 1}, 3217 (1970).
\bibitem{hpy} H. P. Yuen, Phys. Rev. A {\bf 13}, 2226 (1976).
\bibitem{tomb} {\em Squeezed and Nonclassical Light}, P. Tombesi and
E. R. Pike, Eds., NATO Series B {\bf 190}, Plenum Press, New York, 1989.
\bibitem{spe} Special issues on Squeezed states:
J. Opt. Soc. Am. B{\bf 4},
(H. J. Kimble and D. F. Walls, Eds.,1987); J. Mod. Opt. {\bf 34}
(R. Loudon
and P. L. Knight, Eds., 1987).
\bibitem{yam} Y. Yamamoto and 
H. A. Haus, Rev. Mod. Phys. {\bf 58}, 1001 (1986).
\bibitem{iss} {\em Quantum Interferometry}, F. De Martini et al, Eds.
(VCH, Wenheim 1996), pp. 95-224.
\bibitem{schiller} G. Breitenbach, S. Schiller and J. Mlynek, 
Nature {\bf 387}, 471 (1997).
\bibitem{yaim} Y. Yamamoto, N. Imoto, and S. Machida, Phys. Rev. A
{\bf 33}, 3243 (1986).
\bibitem{banana} Y. Yamamoto, S. Machida, N. Imoto, M. Kitagawa, and
G. Bj\"ork, J. Opt. Soc. Am. B {\bf 4}, 1645 (1987).
\bibitem{arrow} E. B. Davies, {\em Quantum Theory of Open Systems},
Academic Press, New York, 1973.
\bibitem{randomlaser} T. J. Jaseja, A. Javan, and C. H. Townes,
Phys. Rev. Lett. {\bf 10}, 165 (1963).
\bibitem{sg} L. Susskind and J. Glogower, Physics {\bf 1}, 49 (1964).
\bibitem{shsh} J. H. Shapiro and S. R. Shepard, Phys. Rev. A {\bf 43}, 
3795 (1991).
\bibitem{ban} M. Ban, Phys. Rev. A {\bf 50}, 2785 (1994); 
Special issue on {\em Quantum Phase and Phase Dependent Measurements}, 
Physica Scripta T {\bf 48} (1993).
\bibitem{stima1} A. S. Holevo, {\em Probabilistic and statistical
aspects of quantum theory}, North Holland, Amsterdam, 1982.
\bibitem{stima2} C. W. Helstrom, {\em Quantum detection and estimation
theory}, Academic Press, New York, 1976.
\bibitem{dms} G. M. D'Ariano, C. Macchiavello and M. F. Sacchi,
Physics Letters A, {\bf 248} 103 (1998)
\bibitem{notabene} The set of phase pure states indeed has to be
restricted excluding the states $\hat\varrho $ with $\varrho _{ij}\neq
0$ only for $i-j=nk$, with $n\in \mathbb N$ and $k$ denoting an integer 
constant $\geq 2$. In fact, those states have phase properties that
are periodic of $2\pi/k$. A simple example is given by the 
superposition of two coherent states with amplitude $\pm \alpha$ 
(Schr\"odinger-cat like states), for which $k=2$.
\bibitem{lind} G. Lindblad, Commun. Math. Phys {\bf 48}, 199 (1976).
\bibitem{holevo} A. S. Holevo, J. Funct. Anal. {\bf 131}, 255 (1995).
\bibitem{peres} A. Peres, {\em Quantum Theory: Concepts and Methods}, 
Kluwer Academic Publishers, Dordrecht, 1998.
\end{thebibliography}
\end{document}